\documentclass[12pt]{article}

\usepackage{epsfig}
\usepackage{amssymb}
\usepackage{subfigure}
\usepackage{graphicx}
\usepackage{cite}

\usepackage[dvipsnames]{xcolor}

\usepackage{jheppub} 

\begin{document}

\baselineskip 6mm
\renewcommand{\thefootnote}{\fnsymbol{footnote}}


\newcommand{\nc}{\newcommand}
\newcommand{\rnc}{\renewcommand}



\newcommand{\tcb}{\textcolor{blue}}
\newcommand{\tcr}{\textcolor{red}}
\newcommand{\tcg}{\textcolor{green}}


\def\ba{\begin{array}}
\def\ea{\end{array}}
\def\be{\begin{eqnarray}}
\def\ee{\end{eqnarray}}
\def\bea{\begin{eqnarray}}
\def\eea{\end{eqnarray}}
\def\nn{\nonumber\\}


\def\ct{\cite}
\def\la{\label}
\def\eq#1{\eqref{#1}}


\def\a{\alpha}
\def\b{\beta}
\def\g{\gamma}
\def\G{\Gamma}
\def\d{\delta}
\def\D{\Delta}
\def\e{\epsilon}
\def\et{\eta}
\def\ph{\phi}
\def\Ph{\Phi}
\def\ps{\psi}
\def\Ps{\Psi}
\def\k{\kappa}
\def\l{\lambda}
\def\L{\Lambda}
\def\m{\mu}
\def\n{\nu}
\def\th{\theta}
\def\Th{\Theta}
\def\r{\rho}
\def\s{\sigma}
\def\S{\Sigma}
\def\ta{\tau}
\def\o{\omega}
\def\O{\Omega}
\def\pr{\prime}


\def\half{\frac{1}{2}}
\def\goto{\rightarrow}

\def\na{\nabla}
\def\grad{\nabla}
\def\curl{\nabla\times}
\def\div{\nabla\cdot}
\def\pa{\partial}
\def\fr{\frac}

\def\bra{\left\langle}
\def\ket{\right\rangle}
\def\lb{\left[}
\def\lc{\left\{}
\def\ls{\left(}
\def\lp{\left.}
\def\rp{\right.}
\def\rb{\right]}
\def\rc{\right\}}
\def\rs{\right)}

\def\vac#1{\mid #1 \rangle}


\def\td#1{\tilde{#1}}
\def\check{ \maltese {\bf Check!}}


\def\Tr{{\rm Tr}\,}
\def\det{{\rm det}}
\def\text#1{{\rm #1}}


\def\bc#1{\nnindent {\bf $\bullet$ #1} \\ }
\def\ch {$<Check!>$ }
\def\ss {\vspace{1.5cm}}
\def\inf{\infty}

\begin{titlepage}

\hfill\parbox{5cm} { }

 
\vspace{25mm}

\begin{center}
{\Large \bf Finite size effect on quantum correlations}

\vskip 1. cm
   {Chanyong Park\footnote{e-mail : cyong21@gist.ac.kr} and Jung Hun Lee\footnote{e-mail : junghun.lee@gist.ac.kr}}

\vskip 0.5cm

{\it  Department of Physics and Photon Science, Gwangju Institute of Science and Technology,  Gwangju  61005, Korea}

\end{center}

\thispagestyle{empty}

\vskip2cm


\centerline{\bf ABSTRACT} \vskip 4mm

\vspace{0.5cm}

We holographically study the finite-size scaling effects on macroscopic and microscopic quantum correlations deformed by excitation and condensation. The excitation (condensation) increases (decreases) the entanglement entropy of the system. We also investigate the two-point correlation function of local operators by calculating the geodesic length connecting two local operators. As opposed to the entanglement entropy case, the excitation (condensation) decreases (increases) the two-point function. This is because the screening effect becomes strong in the background with the large entanglement entropy. We further show that the holographic renormalization leads to the qualitatively same two-point function as the one obtained from the geodesic length.

\vspace{2cm}

\end{titlepage}





\section{Introduction\label{sec:1}}

Based on the AdS/CFT correspondence or holography \cite{Maldacena:1997re,Witten:1998qj,Witten:1998zw,Gubser:2002tv}, a variety of holographic studies have been performed for understanding nonperturbative aspects of strongly interacting systems. There is also a growing interest in understanding more fundamental structures and features of many quantum systems, like quantum entanglement \cite{Srednicki:1993im,Ryu:2006bv,Ryu:2006ef,Hubeny:2007xt,Nozaki:2012zj,Bhattacharya:2012mi,MIyaji:2015mia,Rangamani:2016dms,Takayanagi:2017knl} and complexity \cite{Brown:2015bva,Brown:2015lvg,Takayanagi:2018pml,Goto:2018iay,Bernamonti:2019zyy,Bernamonti_2020}. The recent holographic studies on entanglement entropy provide a new fascinating tool to figure out nonperturbative features of quantum entanglement from the geometrical point of view. For a critical system described by a conformal field theory (CFT), intriguingly, it was shown that a minimal surface extending to a holographic dual geometry can reproduce the same as the result obtained in a CFT \cite{Casini:2004bw,Ryu:2006bv,Ryu:2006ef,Casini_2007,Albash:2011nq,Klebanov:2012yf,Cremonini:2013ipa,Faulkner:2014jva,Park:2014gja,Park:2015hcz,Bueno:2016rma,Jang:2017gwd,Park:2018ebm,Park:2019pzo}. In this work, we consider a CFT deformed by a relevant operator and investigate macroscopic and microscopic quantum correlations in this deformed theory. 
 
According to the AdS/CFT correspondence, the radial coordinate of an AdS space can be regarded as the energy scale of a dual quantum field theory (QFT) \cite{Witten:1998qj,Witten:1998zw,Gubser:2002tv,deBoer:1999tgo}. Intriguingly, it was shown that the Hamilton-Jacobi equation of the gravity theory can match to the renormalization group (RG) flow equation of the dual QFT. In this case, since the RG flow is parameterized by the energy scale, it corresponds to the momentum space RG flow usually used in a QFT. On the other hand, many peoples recently have paid attention to the quantum entanglement entropy for understanding various quantum aspects related to quantum information. Based on the AdS/CFT correspondence, Ryu and Takayanagi proposed a new conjecture that the entanglement entropy of a QFT can be evaluated by calculating the area of the minimal surface extending to the dual geometry \cite{Ryu:2006bv}. In this procedure, the subsystem size is reinterpreted as the inverse of the energy scale, which is similar to the real space RG flow usually used in the condensed matter theory \cite{Vidal:2007hda,Albash:2011nq,Myers:2012ed,Swingle:2009bg,MolinaVilaplana:2011jy,Nozaki:2012zj,Okunishi:2012yh,Matsueda:2012gc,Kim:2014qpa,Kim:2014yca,Park:2015afa}. In the holographic setup, therefore, we have two different descriptions for the RG flow, the momentum \cite{deBoer:1999tgo,Balasubramanian:1999jd,deBoer:2000cz,Bianchi:2001kw,Skenderis:2002wp} and real space RG flows \cite{Kim:2016hig,Kim:2016jwu,Park:2018ebm,Park:2019pzo,Park:2018snf}. They may be associated with each other by the Fourier transformation.

In Ref. \cite{Calabrese:2004eu}, the authors showed that, when a CFT is deformed by a relevant operator, the entanglement entropy can have nontrivial corrections associated with the subsystem size. From the real space RG flow viewpoint, this finite-size effect is crucially related to the effect of the RG flow. In general, the RG flow depends on what kind of deformation changes a UV CFT. In this work, we consider two different types of deformation like excitation and condensation. On the dual gravity side, the excitation is associated with the black hole geometry, whereas the condensation is described by the bulk matter fields. These two different deformations may give rise to a different effect on the entanglement entropy. The reason is that the excitation usually increases the degrees of freedom, while the condensation reduces the degrees of freedom of the underlying theory. Due to this reason, the entanglement entropy with the excitation or condensation increases or decreases along the RG flow in the UV region, respectively. We clearly show this feature by calculating holographically the entanglement entropy of the deformed theory.    

The entanglement entropy can be regarded as a macroscopic quantum correlation because it describes the quantum correlation between two macroscopic subsystems. We can also take into account a two-point correlation function of two local operators which corresponds to a microscopic quantum correlation. According to the AdS/CFT conjecture, these two different macroscopic and microscopic quantum correlations can be described by a similar geometric object on the dual gravity side. More precisely, if we concentrate on a two-dimensional QFT, the entanglement entropy and two-point function are described by a one-dimensional geodesic curve connecting two boundary points on the dual gravity side \cite{Susskind:aa}. In this case, two boundary points correspond to two boundaries of the entangling region for the entanglement entropy and the positions of two local operators for the two-point function. Intriguingly, we show by explicit holographic calculation how the macroscopic and microscopic correlations are related to each other for a two-dimensional QFT. In this case, we crucially exploit the conjecture that explains how the two-point function is realized by the geodesic curve \cite{Susskind:aa}. To check the validity of this conjecture, we further investigate the two-point function of two local operators by using the holographic renormalization technique \cite{Balasubramanian:1999jd,deBoer:2000cz,DeWolfe:2000xi,Bianchi:2001kw,Skenderis:2002wp} which leads to the same as the qualitative feature expected in the conjecture.

The rest of this paper is organized as follows: In section 2, we review the basic setup for computing both the finite-size effect of the entanglement entropy and the associating correlation function in the deformed backgrounds. In section 3, applying the RT formulation to several deformed backgrounds, we survey quantum corrections of the entanglement entropy due to the finite size effect and also compute two-point functions, which can be constructed by exponentiating the bulk geodesics anchored at the boundary entangling points, in both a UV and an IR limits. In section 4, we study the two point-functions by using the holographic renormalization method and compare the asymptotic behaviors of the correlation functions with the results obtained in section 3. In section 5, we consider the theory with mixed deformations containing excitation and condensation and survey the entanglement entropy and two-point function.
Finally, we close this work with some concluding remarks in section 6.


\section{Finite size corrections}
\label{sec:2}

Recently, a variety of strongly interacting conformal field theories (CFT) has been widely studied by using the AdS/CFT correspondence. When the conformal symmetry defined at a UV fixed point breaks down due to deformations, the UV CFT usually evolves to a nontrivial effective field theory as the observation energy scale becomes lower. To understand this procedure clearly, it would be important to figure out the non-perturbative renormalization group flow. Despite the importance of the non-perturbative feature of the RG flow, unfortunately, we have no well-established mathematical and physical method. In this situation, the AdS/CFT correspondence and its generalization may be helpful to investigate the underlying structure of the non-perturbative RG flow. In this work,  we will discuss how microscopic and macroscopic correlations of quantum systems change along the RG flow by performing the holographic entanglement entropy calculation \cite{Calabrese:2004eu,Calabrese:2009qy}.

Before studying the RG flow of quantum correlations, it is worth noting that the RG flow of the entanglement entropy is different from the RG flow studied in the holographic renormalization procedure. From the dual field theory viewpoint \cite{Ryu:2006bv,Ryu:2006ef}, more precisely, the entanglement entropy RG flow represents the real space RG flow, while the holographic renormalization studied in Ref. \cite{deBoer:1999tgo} corresponds to the momentum space RG flow. In the holoraphic renormalization, the AdS boundary can move and its position is identified with the energy scale of the dual field theory. On the other hand, the energy scale of the dual field theory in the entanglement entropy RG flow is given by the inverse of the subsystem size. Therefore, the RG effect in the real space RG flow is expressed as the finite-size effect with the effect.

Now, let us assume a two-dimensional CFT at a UV fixed point. When the UV CFT is deformed by a certain dimensionful operator, the entanglement entropy is modified. Dividing a system into a subsystem with a size $l$ and its complement, the resulting entanglement entropy in the UV region may have the following general expansion form
\be			\la{ansatz:EE}
S_E = \frac{c}{3} \log \fr{l}{\e} + {\rm constant} + \frac{c}{3}  \ S_{F} ,
\ee 
where $\e$ and $c$ denote an appropriate UV cutoff and a central charge of the UV CFT, respectively. Above the first two terms are the entanglement entropy of the UV CFT, whereas the last term $S_{F} $ was called the finite size effect \cite{Calabrese:2004eu} which is associated with the deformation breaking the conformal symmetry. In the real space RG flow described by the entanglement entropy the inverse of the subsystem size plays the role of the energy scale observing the system, so that we can reinterpret $l$ as the inverse of the RG scale. Since the finite size effect disappears for an undeformed critical system or at a UV fixed point with $l \to 0$, the above entanglement entropy automatically reduces to the well-known result of a two-dimensional CFT 
\be			\la{Result:EE}
S_E = \frac{c}{3} \log \fr{l}{\e} + {\rm constant} .
\ee 
When the theory deforms near the UV fixed point, the finite size effect is given by a function of a dimensionless variable. Introducing an appropriate correlation length as a dimensionful parameter, the finite size effect can be is given by a function of $l$ and $\xi$ like $S_{F} \ls l/\xi \rs$. In this case, the correlation length characterizes the effect of the deformation.

Since the correlation length usually depends on the microscopic details at a given energy scale, it is worth noting that the correlation length can varies as the RG scale changes. Then, what is the possible form of the finite size effect? Although it is interesting and important to answer this question, it is not easy because it requires a full non-perturbative analysis. However, the AdS/CFT correspondence can give us a chance to understand the form of the non-pertubative finite size effect in several specific regions. For example, the UV region deviated slightly from the UV fixed point can be characterized by $l \ll \xi $ in the real space RG flow. In this case, a dimension counting with a small expansion parameter $l/\xi $ enables us to guess the following general expansion form 
\be			\la{Result:UVexpansion}
S_{F}  = \sum_{j \ge 1} \  s_j   \ls \fr{l}{\xi} \rs^{2 j \a } ,
\ee 
where $s_j$ and $\a$ are two constants. Here the factor $2$ in the expansion power appears because we assume that the underlying theory is invariant under $l \to - l$. Although this expansion looks universal in the UV region, determining the exact values of $s_j$ and $\a$ remains an important issue to understand the RG flow effect of the deformation. Intriguingly, the similar structure may occur in other physical quantity like a two-point function because its finite size effect is also characterized by the same correlation length. In the next sections, we will clarify that $\a$ is related to the conformal dimension of a deformation operator.

In an IR region, on the other hand, the correlation length usually has a different value from the UV one because the strength of interactions can vary along the RG flow. For a two-dimensional CFT deformed by an effective mass $m$, the correlation length in the IR regime is given by as $\xi  = 1/m$. In this case, the IR region is specified by $\xi /l \ll 1$ and the IR finite size effect may be expressed alternatively by the following expansion form
\be			\la{res:IRentanglemententropy}
S_{F}   =  \log  \fr{\xi }{l} + \sum_{j \ge 1}  \ s_j^{\infty}  \ls  \fr{ \xi}{l} \rs^{2 j \b} .
\ee
where $s_j^{\infty}$ and $\b$ are another constants defined in the IR region. In the ultimate IR limit with $\xi /l \to 0$, the entanglement entropy results in
\be			\la{result:IRee}
S_E = \fr{c}{3} \log \fr{\xi }{\e} ,
\ee
which is the well-known entanglement entropy of a noncritical QFT in the IR limit. In the holographic setup described in a three-dimensional dual AdS space, the above known entanglement entropies in \eq{Result:EE} and \eq{result:IRee} were reproduced \cite{Ryu:2006bv,Ryu:2006ef}. However, if we consider a different deformation, the finite size effect may lead to a totally different form in the IR region due to the nontrivial RG flow.

\subsection{Finite size effect on a microscopic two-point function}

As mentioned before, the RG effect of a deformed theory can be described by a finite size effect. This is also true for a two-point correlation function because the deformation is characterized by the same correlation length. In other words, a two-point correlation function of a local operator can be written as the following general form
\be
\bra O (x_1^\m) \, O( x_2^\m) \ket  \sim  \fr{e^{- F\ls |x_1^\m-x_2^\m|/ \xi \rs }}{|x_1^\m-x_2^\m|^{2 \D_O}}  ,
\ee
where $\D_O$ is a conformal dimension of the operator $O$ and $x_i^\m=\lc t_i, x_i \rc$ indicates the position of a local operator. In general, $F$ is a highly nontrivial function relying on the proper distance of two operators, $|x_1^\m-x_2^\m| = \sqrt{ - (t_1-t_2)^2 + (x_1 -x_2)^2}$ in a two-dimensional Minkowskian space. In the CFT case, the correlation length $\xi$ diverges due to the conformal symmetry and the function $F$ vanishes. Finally, the above general two-point function reduces to the well-known two-point function of a CFT. When a CFT deforms, the deformation leads to a nontrivial function $F$ with the correlation length relying on the RG scale. This fact implies that the function $F$ crucially relies on the RG flow caused by the deformation. In the UV region, there exists a small parameter, $|x_1^\m-x_2^\m|/\xi \ll 1$. Expanding the general two-point function in terms of this small parameter, the resulting two-point function is perturbatively given by
\be         \la{result:tpointfn}
\bra O (x_1^\m) \, O( x_2^\m)\ket  \sim  \fr{1}{|x_1^\m-x_2^\m|^{2 \D_O}} \lb 1 + \sum_{i=1}^{\inf} c_i \ls \fr{|x_1^\m-x_2^\m|}{\xi} \rs^{2 i \a}  \, \rb .
\ee
where $\a$ is the constant used in the UV expansion of the entanglement entropy. 

If we concentrate on the equal-time correlation function satisfying $t_1-t_2=0$, the resulting two-point function is rewritten as
\be         \la{Result:equaltimefunction}
\bra O (x_1) \, O( x_2)\ket  \sim  \fr{1}{|x_1 -x_2|^{2 \D_O}} \lb 1 + \sum_{i=1}^{\inf} c_i \ls \fr{|x_1- x_2|}{\xi} \rs^{2 i \a}  \, \rb ,
\ee
where $|x_1 -x_2|$ indicates a distance in the $x$-direction. Intriguingly, replacing $|x_1-x_2|$ with $l$ in the previous section, the finite size effect of the two-point function gives rise to the same expansion as that of the entanglement entropy up to an overall factor. This is because the expansion form depends mainly on the deformation. Therefore, the same $\a$
corresponding to the conformal dimension of the deformation appears in both expansions of the entanglement entropy and two-point function. This fact becomes more manifest in the dual gravity model.

In the holographic study, a physical quantity of the field theory can be realized by a geometric object on the dual gravity side \cite{Susskind:aa}. The entanglement entropy studied in the previous section is described by a minimal surface area extending to the dual geometry. Similarly, it was proposed that a two-point function at a given time can be represented as a geodesic curve connecting two local operators via
\be    		\la{Formula:two-pt function}
\bra O (x_1) \, O (x_2) \ket \sim e^{-m L(|x_1-x_2|)}  ,
\ee
where $L$ indicates a geodesic length of a particle with a mass $m$. For a three-dimensional gravity theory, interestingly, a minimal surface describing the entanglement entropy is also given by a geodesic curve connecting two entangling points. As a result, the entanglement entropy and two-point function are described by the same curve in the holographic setup. This fact shows that at least for a two-dimensional field theory, the macroscopic entanglement entropy is closely associated with the microscopic two-point function. Hereafter, we discuss the validity of the proposed formula in \eq{Formula:two-pt function} and investigate the relation between the entanglement entropy and two-point function with a variety of deformation.

Before closing this section, we discuss how the proposed formula \eq{Formula:two-pt function} reproduces the known two-point function of a two-dimensional CFT. To describe a local scalar operator holographically, we consider a bulk massive scalar field living in a three-dimensional AdS space  
\be
ds^2 = \fr{R^2}{z^2}  \ls - dt^2 + dx^2 + dz^2\rs ,
\ee
where $R$ indicates an AdS radius. In this setup, the fixed background geometry implies that we take into account a probe limit where the gravitational backreaction of the bulk scalar field is ignored. According to the AdS/CFT correspondence, the bulk massive scalar field maps to a scalar operator of the dual QFT. More precisely, the bulk massive scalar field is governed by the following action
\be
S = \int d^{3} x  \sqrt{-g} \, \ls   - \half \pa_M \ph \, \pa^M \ph - \half  M^2 \ph^2 \rs ,
\ee
When the mass of the bulk scalar field satisfies 
\be			\la{Result:scalarmass}
M^2 = - \fr{\D_O  \ls 2 - \D_O \rs}{R^2} < 0 ,
\ee
it maps to the scalar operator $O (x)$ with a conformal dimension $\D_O$ on the dual field theory side. Using this relation together with the previous two-point function in \eq{Formula:two-pt function}, it is possible to rederive the two-point function of a scalar operator holographically. In this procedure, if we relate the bulk field mass $M$ to the particle mass $m$ like
\be
m = \sqrt{- \fr{ \D_O }{(2 - \D_O)} M^2} .
\ee
the expected two-point function of CFT is reproduced from \eq{Formula:two-pt function}. To check this, we calculate the geodesic length when two local operators are located at different positions at equal time. In this case, the geodesic curve on the dual gravity side is given by a curve connecting two local operators. This gravity configuration leads to a equal-time two point function from the dual field theory viewpoint. In this case, the geodesic length is determined by
\be
 L(|x_1-x_2|)  = \int_{x_1}^{x_2} dx \fr{R}{z} \sqrt{1+ z'^2} .
\ee
Note that, if we replace $x_1$ and $x_2$ with $-l/2$ and $l/2$, the geodesic length calculation is the same as the previous entanglement entropy calculation. Finally, the resulting two-point function reads
\be
\bra O(x_1)  \,  O(x_2) \ket  \sim \fr{1}{|x_1-x_2|^{2 \D_O}} ,
\ee
which is the expected form in CFT.

\section{Finite size effect on a two-point function}

In the previous section, we discussed the the perturbative expansions of the entanglement entropy and two-point correlation function. For a two-dimensional deformed field theory, we claimed that these two quantities at least in the UV region lead to a similar expansion with the same expansion power $\a$. To clarify the physical meaning of $\a$, in this section, we will consider a two-dimensional field theory and three-dimensional holographic duals deformed by a relevant operator. Then, we will show that the expansion power $\a$ in the UV region is closely related to the conformal dimension of the deformation operator.

To specify the finite size effect in the UV region, let us first discuss a finite effect of an Euclidean two-dimensional CFT deformed by a mass 
\be
S=  \fr{1}{2} \int d^{2} x  \ls \pa_\m \ph \pa^\m \ph  - m_{eff}^2 \ph^2  \rs ,
\ee
where $m_{eff}$ indicates an effective mass containing all quantum corrections. For the massless case with $m_{eff}=0$, the conformal symmetry is restored. For the mass deformed theory the conformal symmetry is slightly broken in the UV region, so that the effective mass $m_{eff}$ is small. In this case, the conformal dimensions of $\ph$ and $m_{eff}$ are given by $\D_{\ph}=0$ and $\D_m=1$ respectively.

Now, let us consider a two-point function of the filed $\ph$. The two-point function of the mass deformed theory must satisfy the following equation of motion
\be
\lb  -  \pa^2  - m_{eff}^2 \rb   \bra \ph (x_1^{\m}) \,  \ph (x_2^{\m}) \ket  =  \d (x_1^{\m}-x_2^{\m})
\ee
Using the Fourier transformation, the resulting two-point function is given by  
\be
\bra  \ph (x_1^{\m}) \,  \ph (x_2^{\m}) \ket &\sim & 
\int   d^{2} p \  \fr{e^{ - i p_\m (x_1^{\m}-x_2^{\m})} }{ p^{2} - m_{eff}^2}  ,
\ee
where $p^2 = p^\m  p_\m$ for the Euclidean space. In the UV region with the small effective mass, the two-point function is perturbatively given by 
\be
\bra  \ph (x_1^{\m}) \,  \ph (x_2^{\m}) \ket \sim   1  + \sum_{i=1}^{\inf} A_i  \  \ls m_{eff} \, |x_1^{\m}-x_2^{\m}|  \rs^{2 i } ,
\ee
where $A_i$ are nontrivial constants. This result shows that, as we expected, the two-point function reduces to that of CFT in the massless limit. Moreover, the expansion power is given by the conformal dimension of the mass deformation, $\D_m=1$, as claimed before. The correlation length of this mass deformed theory is given by the inverse of the effective mass, $\xi \sim 1/m_{eff}$. The resulting two-point function is coincident with the previous result in \eq{result:tpointfn} with $\a=1$. This is also true for holographic models. In the following sections, we will further  investigate the finite size of two holographic model deformed by different deformation operators.

\subsection{Finite size effect of excitation}
\label{sec:31}

Now, let us take into account another deformation caused by excitation of a massless boson. On the dual gravity, such excitation in the UV region can be described by a AdS black hole. For a three-dimensional case, an AdS black hole is described by  
\be \label{btz}
ds^2 =  \fr{R^2}{z^2} \ls - f(z) dt^2 + dx^2 + \fr{1}{f(z)} dz^2 \rs ,
\ee
with a blackening factor 
\be
f(z)  = 1 - \fr{z^2}{z_h^2} ,
\ee
where $z_h$ denotes a black hole horizon. The black hole geometry usually satisfies the thermodynamics law, which in the holographic setup is reinterpreted as thermodynamics of the dual field theory. From the black hole geometry, the Hawking temperature and Bekenstein-Hawking entropy read
\be
T_H = \fr{1}{2 \pi z_h} \quad {\rm and} \quad S_{BH} = \fr{R}{4G} \fr{L}{z_h} ,
\ee
where $L$ is an appropriately regularized spatial volume of the boundary space-time. Then, the thermodynamics law yields the following internal energy and pressure
\be
E &= &  \fr{L}{16 \pi G} \fr{R}{z_h^2}    , \nn
P &=&   \fr{1}{16 \pi G} \fr{R}{z_h^2}  .
\ee
These two quantities show that the excitation in the dual QFT has an equation of state parameter $w=1$. For the two-dimensional CFT, this corresponds to the equation of state parameter of a massless boson. For a general $d$-dimension, the equation of state parameter of a massless boson is given by 
\be
P =  \fr{1}{d-1} \r ,
\ee
where $\r$ indicates the energy density. In the UV region with $l \ll 1/T_H$, the internal energy is regarded as the excitation energy of a massless boson.

Now we calculate the entanglement entropy on this background. To apply the holographic technique, we first consider a space-like hypersurface at a given time on which the minimal surface is define. At a given time, the induced metric on the space-like hypersurface is given by
\be
ds^2 =  \fr{R^2}{z^2} \ls dx^2 + \fr{1}{f(z)} dz^2 \rs .
\ee
Assuming that two end points of the entangling region are located at $-l/2$ and $l/2$, the resulting entanglement entropy in the UV region reads
\be
S_E = \fr{c}{3}  \ls \log \fr{l}{\e} + \fr{\pi^2}{6} \fr{l^2}{\xi^2}  + \cdots \rs  ,
\ee
where the correlation length is given by $\xi  = 1 /T_H$.

In a three-dimensional gravity theory, since entanglement entropy and two-point function are described by the same geodesic curve on the dual gravity side, the entanglement entropy is closely related to a two-point function. Therefore, one can easily read a two-point function by applying the previously proposed formula in \eq{Formula:two-pt function}. To do so, we first identify the subsystem size $l$ with the distance of two local operators $|x_1-x_2|$. In this procedure, it must be noted that the entanglement entropy is defined on the space-like surface at a given time. In other words, two ends of the entangling region are measured at the same time, $t_1=t_2$. This fact implies that $|x_1-x_2|$ identified with $l$ must be the distance of two local operators measured at the same time. As a consequence, the resulting two-point function derived from \eq{Formula:two-pt function} corresponds to the equal-time two-point function. Representing the geodesic length in terms of the entanglement entropy
\be
L(|x_1-x_2|)  = 4 G S_E(|x_1-x_2|)  ,
\ee
the equal-time two-point function becomes
\be \label{tpfbtz}
\bra O (x_1)  \, O (x_2)\ket  
\sim \fr{1}{|x_1-x_2|^{2 \D_O}}  \ls 1 - \fr{\pi^2  \, \D_O }{3}  \fr{|x_1-x_2|^2}{\xi ^2}+ \cdots \rs .
\ee
where $\D_O$ indicates the conformal dimension of a local scalar operator $O (x)$. Noting that the excitation energy has a conformal dimension $1$, the obtained two-point function is again consistent with the expected UV expansion form in \eq{Result:equaltimefunction}. From the obtained results, we see that the entanglement entropy increases as the correlation length decreases, while the two-point function decreases. This is because the macroscopic quantum correlation of excitation increases as the entanglement entropy increases. The strong correlation of excitation screens the microscopic quantum correlation between two local operators, so that the two-point function of local operators decreases as the entanglement entropy of excitation increases.

In the IR regime ($l \gg z_h$), the geodesics of the RT surface probe the black hole horizon extending tangentially to it. This leads to the linear thermal correction to the entanglement entropy. So, the entanglement entropy is given by
\be
S_E = \fr{R}{4 G} \ls  \fr{2 \pi l}{\xi } + \log \fr{\xi }{ 2 \pi \e} + \cdots \rs
\ee
Here the first term represents the thermal entropy contained in the subsystem \cite{Solodukhin:2011gn}, whereas the second term is the first quantum correction caused by deformation. Applying \eq{Result:equaltimefunction}, this result leads to the following IR two-point function 
\be         \la{Result:IRtwopointfn}
\bra O (x_1)  \,  O (x_2)\ket \sim  \fr{1}{\xi ^{2 \D_0}}  e^{- 2 \pi \D_O |x_1 -x_2| /\xi  }  .
\ee 
This result shows that the IR the two-point function is suppressed exponentially not by power-law discussed in \eq{res:IRentanglemententropy}. Moreover, as the temperature becomes high and the conformal dimension of a local operator is large, the two-point correlation is suppressed more rapidly as the distance of two local operators increases.

\subsection{Finite size effect of condensation}

In the previous sections, we showed that the UV entanglement entropy and two-point function follow the expected expansion form when the conformal dimension of deformations is one. For more general situations, in this section, we take into account a two-dimensional CFT deformed by a relevant operator with a conformal dimension in the range of $1 \le \D_d < 2$. On thd dual gravity side, the dual geometry is described by the following action
\be    \la{Action:scalar}
S = \fr{1}{16 \pi G} \int d^{3} x \sqrt{-g} \lb  \ls {\cal R} - 2 \L\rs - \half \pa_M \ph  \pa^M \ph -  \half M^2  \ph^2  \rb ,
\ee
where $\L$ denotes a cosmological constant, $\L = - 1/ R^2$ with an AdS radius $R$ and $M$ is a bulk mass of the scalar field. In general, the gravitational backreaction of the scalar field deforms the AdS space and the deformed geometry is determined by  
\be 	\la{eq:Einandscalar}
{\cal R}_{MN} -  \half g_{MN} {\cal R} +  g_{MN} \L &=& T_{MN} ,  \nn
\fr{1}{\sqrt{-g}} \pa_M \ls \sqrt{-g} G^{MN} \pa_N \ph \rs - M^2 \ph &=&  0 ,
\ee
where the energy-momentum tensor of the scalar field is given by
\be
T_{MN} =  \half \pa_M \ph  \pa_N \ph   - \fr{1}{4}  g_{MN}  \ls \pa_P \ph  \pa^P \ph +  M^2 \ph^2 \rs  .
\ee
In this case, the bulk mass $M$ is associated with the conformal dimension of the dual scalar operator
\be         \la{Relation:scalarmass}
M^2 = - \fr{\D_d (2 - \D_d)}{R^2} .
\ee

When we take into account deformation by a relevant scalar operator. Near the boundary of an asymptotic AdS space, the scalar field has usually the following perturbative profile
\be
\ph (z)  = J z^{2 - \D_d} (1 + \cdots )+ D z^{\D_d} (1 + \cdots ) ,
\ee
where the ellipsis indicate higher order corrections. On the dual CFT, $J$ and $D \equiv \bra O\ket$ are identified with the source and the vev of the deformation operator. Note that for $\D_d=1$ the source term can have an additional logarithmic term due to the degeneracy of two independent solutions. From now on, we focus on the case of $J=0$ to study the effect of the deformation operator. Once we set $J=0$, we don't worry about the additional logarithmic term for $\D_d=1$ because the degeneracy disappears. Note that the mass of the dual bulk scalar field must have $-1/R^2 \le M^2 \le 0$ for a relevant deformation.

Assuming that the bulk scalar field relies only on the $z$-coordinate, the most general ansatz compatible with the isometry is given by
\be     \la{Result:condensationmetric}
ds^2 = \fr{R^2}{z^2} \ls g(z) \et_{\m\n} dx^\m dx^\n + dz^2 \fr{}{} \rs ,
\ee 
where $ \et_{\m\n} $ denotes a two-dimensional Minkowskian metric. Recalling that the equations of motion are invariant under $z \to -z$, the solutions must have the following expansion form to satisfy Einstein equations and scalar field equation at the same time 
\be	
\ph (z) &=&  \fr{ z^{\D_d}}{z_d^{\D_d}} \lb 1 +  \sum_{j=1}^{\inf} a_{j} \ \ls \fr{z}{z_d} \rs^{2 j \D_d  }  
  \rb  , \nn
g  (z)  &=& 1 +  \fr{ z^{2 \D_d}}{z_d^{2 \D_d}}  \lb c_{0} +  \sum_{j=1}^{\inf} c_{j} \ \ls \fr{z}{z_d} \rs^{2 j \D_d  }   \rb  ,
\ee
where we set $D = 1/z_d^{\D_d}$ for convenience. After substituting these expansions into the equations of motion, solving them perturbatively gives rise to the following perturbative solution
\be
\ph (z) &=&   \fr{ z^{\D_d}}{z_d^{\D_d}}   \ls  1 +  \fr{\D_d}{8 (2 \D_d - 1)}  \fr{ z^{2\D_d}}{z_d^{2 \D_d}}   + \cdots  \rs  , \nn
g  (z)  &=&  1 -  \fr{ z^{2 \D_d}}{z_d^{2 \D_d}}  \ls \fr{1}{4}   +  \fr{2-\D_d}{64 (2 \D_d -1)}  \fr{ z^{2\D_d}}{z_d^{2 \D_d}}   + \cdots \rs . 
\ee

On this background geometry, the entanglement entropy is given by
\be
S_E = \fr{1}{4 G} \int_{-l/2}^{l/2} dx \fr{R}{z} \sqrt{g (z) + z'^2}
\ee
Solving the equation,
the subsystem size is determined by the turning point $z_0$ and the vev of the operator $z_)$ 
\be
l 
&=& 2 z_0 \lb 1 +  \frac{ \Gamma \left(\Delta_d +\frac{3}{2}\right)-\sqrt{\pi } \Delta_d  \Gamma (\Delta_d +1)  }{ 8 \ \Gamma \left(\Delta_d +\frac{3}{2}\right)} \  \fr{z_0^{2 \Delta_d } }{z_d^{2 \Delta_d}} + {\cal O} \ls  \fr{z_0^{4 \Delta_d } }{z_d^{4 \Delta_d}}  \rs \rb .
\ee
Performing the integral of the entanglement entropy after substituting the obtained solution, the resulting entanglement entropy in the UV region expands into
\be
S_E =  \frac{R}{2 G}  \lb  \log \fr{l}{\e} -\frac{\sqrt{\pi }  \  \Gamma (\Delta_d +1)}{ 2^{2 \Delta_d + 4}  \ \Gamma \left(\Delta_d +\frac{3}{2}\right)}   \ \fr{  l^{2 \Delta_d } }{z_d^{2 \Delta_d }}  + {\cal O} \ls  \fr{l^{4 \Delta_d } }{z_d^{4 \Delta_d}}  \rs  \rb .
\ee
This result shows that the UV expansion of the entanglement entropy follows the expected form in \eq{Result:UVexpansion}. In this case, the correlation length is given by $\xi \sim z_d$. Following the prosed formula in \eq{Formula:two-pt function}, the equal-time two-point correlation function becomes
\be 			\la{Result:condtwoptfun}
\bra O (x_1) \, O (x_2)\ket  
\sim  \fr{1}{|x_1-x_2|^{2 \D_O}} \ls 1  +\frac{ 2 \sqrt{\pi } \, \D_d  \  \Gamma (\Delta_d+1)}{ 2^{2 \Delta_d + 4}  \ \Gamma \left(\Delta_d +\frac{3}{2}\right)}   \ \fr{ |x_1-x_2|^{2 \Delta_d } }{\xi^{2 \Delta_d }} + \cdots  \rs ,
\ee
where $\D_O$ and $\D_d$ indicate the conformal dimensions of the local operator and deformation operators respectively. Unlike the previous excitation case, the sign of the second term of the two-point function is positive. Thus, the vev of the deformation operator decreases the entanglement entropy but increases the microscopic quantum correlation between two local operators. This implies that condensation of the deformation operator reduces the screening effect opposite to the previous excitation case.

\section{Holographic renormalization} 
\label{sec:4}

In the previous section, we considered two-dimensional field theory deformed by excitation or condensation and calculated the equal-time two-point functions by using the geodesic length extending to the dual geometry. In this process, the formula in \eq{Formula:two-pt function} was crucial even though there is no exact proof. In this section, we discuss two-point functions of the same systems by using a different way called the holographic renormalization. This calculation may support the formula \eq{Formula:two-pt function} and the UV expansion of the two-point function in \eq{Result:equaltimefunction}. 

\subsection{Holographic two-point function deformed by excitation }
\label{sec:41}

We first study the excitation effect on the two-point function by using the holographic renormalization technique \cite{Gubser:1998bc,Freedman:1998tz,Balasubramanian:1998de,LIN2019114728}. To do so, we consider a scalar field fluctuation living in a three-dimensional AdS black hole which, as mentioned before, describes excitation of massless bosons on the dual field theory side. Assuming that the scalar field fluctuation has the mass satisfying \eq{Result:scalarmass}, the Fourier mode of the scalar field
\be
\ph(z, x^\m) = \int d^2 k \ \ph(z,k ) e^{i k_\m x^\m} ,
\ee 
is governed by
\bea
\phi''(z,k^\m)-\frac{1}{z} \phi'(z,k^\m)+\biggl(\frac{\Delta_O (2-\Delta_O)}{z^2}-k^2+\frac{\Delta_O (2-\Delta_O)}{z_h^2}\biggr)\phi(z,k^\m)&=&0,  \label{PKG} 
\eea
where $k^2 =  k^\m k_\m$ and we ignore higher-order terms. The solution of this equation is given by the following analytic functions
\bea
\phi(z,k^\m) &=&c_1\,\phi_{1}(z,k^\m)+c_2\,\phi_{2}(z,k^\m) \nonumber \\
&=&c_1\, z\, K_{\Delta_O-1}\biggl(z\sqrt{k^2-\frac{(2-\Delta_O)\Delta_O}{z_h^2}}\biggr)
+c_2\, z\, I_{\Delta_O-1}\biggl(z\sqrt{k^2-\frac{(2-\Delta_O)\Delta_O}{z_h^2}}\biggr)
\eea
where $K_\nu(\cdot)$ and $I_\nu(\cdot)$ are the modified Bessel functions of the second and  first kind, respectively. In this case, the bulk-to bulk propagator between scalar fields located at $z$ and $z'$ is defined by
\be         \la{Formula:bulktobulk}
G(z,z';k^\m)=\frac{\theta(z-z')\phi_1(z,k^\m)\phi_2(z',k^\m)+\theta(z'-z)\phi_1(z',k^\m)\phi_2(z,k^\m)}{\sqrt{\vert\gamma\vert}(\phi_1(z',k^\m)\phi'_2(z',k^\m)-\phi'_1(z',k^\m)\phi_2(z',k^\m))}.
\ee
where $\vert\gamma\vert=g^{zz}\vert g\vert$ and $\th$ indicates a step function.

After taking the $z' \to 0$ limit, the boundary-to-bulk propagator $G(z,0;x_1,x_2)$ in the Euclidean space is given by the Fourier transformation of $G(z,0;k^\m)$ 
\bea				\la{Result:htwopoint}
G(z,0;x_1^\m,x_2^\m) 
&=&\frac{1}{2\pi}\int^\infty_0 dk \,k \,J_0(k\vert x_1^\m-x_2^\m \vert) \, G(z,0;k^\m), \label{kzxx}
\eea
with $G(z,0;k^\m)$ with an appropriate normalization   
\bea \label{NGreen}
G(z,0;k)=\frac{( k^2 z_h^2- (2-\Delta_O)\Delta_O)^{(\Delta_O-1/2)}}{2^{\Delta_O-1}\Gamma(\Delta_O) z_h^{\Delta_O-1}}\times \phi_1(z,k) ,  \label{bbp}
\eea
where $k = | k^\m|$ in the Euclidean space and $J_0 (\cdot)$ indicates the Bessel function of the first kind. After performing the integral \eq{Result:htwopoint} and taking the $z\to 0$ limit, the inverse Wick rotation finally gives rise to a Minkowskian two-point function of the dual scalar operator (see Appendix for more details)
\bea \label{slbtb}
\bra O (x_1^\m) \, O (x_2^\m) \ket \sim \frac{1}{\vert x_1^\m-x_2^\m \vert^{2\Delta_O}}\biggl(1-\frac{ \pi^2 (2-\Delta_O)\Delta_O}{16 (\Delta_O-1)}\frac{\vert x_1^\m - x_2^\m\vert^2}{\xi^2}
+\cdots\biggr)  ,
\eea 
where $\xi = 2 \pi z_h = 1/T_H$ and we assume $\vert x_1^\m - x_2^\m\vert^2/{z_h} \ll 1$ which corresponds to the UV region. If we further consider an equal time correlation, $\vert x_1^\m-x_2^\m \vert$ reduces to $\vert x_1-x_2\vert$. Consequently the resulting equal time two-point function in the UV region has the same form as the result in \eq{tpfbtz}, although the coefficient of the first correction is slightly different from the previous result.

Now, we consider another limit satisfying $\vert x_1^\m - x_2^\m\vert^2/{z_h} \gg 1$ which corresponds to the IR region. When performing the integral \eq{Result:htwopoint} for  $\vert  x_1^\m - x_2^\m \vert^2/{z_h} \gg 1$, it finally leads to the following equal time two-point function
\bea \label{short1}
\bra O (x_1) \, O (x_2) \ket \sim \frac{1}{z_h^{2\Delta_O}}e^{- \sqrt{(2-\Delta_O)\Delta_O} \vert x_1 - x_2 \vert/ z_h } \ls 1 + \cdots \rs,
\eea
where the ellipsis implies higher order corrections. In the IR region, this holographic renormalization result is also the same form as \eq{Result:IRtwopointfn} which suppresses exponentially as the distance of two local operators increases.

\subsection{Holographic two-point function deformed by condensation }

Now, we repeat the similar calculation in the dual geometry deformed by condensation of a deformation operator. We first assume that a two-dimensional UV CFT deforms by operator's condensation with a conformal dimension $\D_d$. The dual geometry of this deformed theory is given by \eq{Result:condensationmetric}. On this condensation geometry, the equation of motion for a scalar field fluctuation is governed by
\be
\phi''(z,k^\m)-\frac{1}{z} \phi'(z,k^\m)+\biggl(\frac{\Delta_O (2-\Delta_O)}{z^2}-k^2+\frac{\Delta_O (2-\Delta_O) z^{2\D_d - 2}}{z_d^{2 \D_d}}\biggr)\phi(z,k^\m)&=&0,  
\ee
where we again assume that the mass of $\ph$ is given by \eq{Relation:scalarmass} and the vev of the deformation operator is replaced by $D=1/z_d^{\D_d}$. By using the previous bulk-to-bulk propagator formula \eq{Formula:bulktobulk}, the equal time two-point function in the UV region finally results in
\bea   \label{tpfd}         
\bra O (x_1 ) \, O (x_2) \ket \sim \frac{1}{\vert x_1-x_2 \vert^{2\Delta_O}}\biggl(1 
+ C_{\Delta_O} 
\frac{\vert x_1 - x_2 \vert^{2 \D_d}}{\xi^{2 \D_d}}
+\cdots\biggr)  ,
\eea
where the correlation length is given by $\xi=z_d$. In the case of $\Delta_O=3/2$ and $\Delta_d=3/2$, for example, the coefficient of the first correction in \eqref{tpfd} becomes $C_{\Delta_O}\approx 0.0024$. This result supports  that the two-point function \eq{Result:condtwoptfun} derived from the geodesic length gives rise to the consistent result, as we expected.

\section{Competition of two different deformations}

Now, we consider a two-dimensional field theory containing excitation and condensation discussed before.
A most general metric ansatz of this system is given by
\be
ds^2 = \fr{R^2}{z^2} \ls - \td{f}(z)  \, g (z)  \, dt^2 + g  (z)  \, dx^2 + \fr{1}{\td{f}(z)}  \,dz^2 \rs .
\ee
To describe excitation, we need a black hole solution which usually breaks the boundary Lorentz symmetry. In the above metric ansatz, therefore, $g_{tt}$ and $g_{xx}$ have different metric factor. Ignoring the source term for condensation as done in the previous section, the perturbative expansion forms of bulk fields in the asymptotic region are given by
\be		\la{Metric:defBH}
\ph (z) &=&  \fr{ z^{\D_d}}{z_d^{\D_d}} \lb 1 +  \sum_{j=1}^{\inf} a_{0,j} \ \ls \fr{z}{z_d} \rs^{2 j \D_d }  
+ \sum_{i=1}^\inf  \sum_{j=1}^{\inf} a_{i,j} \  \ls \fr{z }{z_h } \rs^ {2 i}  \ls \fr{z}{z_d} \rs^{2 j \D_d }     \rb  , \nn
\td{f} (z)  &=& 1 + \fr{z^2}{z_h^2}  \lb b_{0,0} +  \sum_{i=1}^\inf  \sum_{j=1}^{\inf} b_{i,j} \  \ls \fr{z }{z_h } \rs^ {2 i}  \ls \fr{z}{z_d} \rs^{2 j \D_d }  \rb  ,\nn 
g  (z)  &=& 1 +  \fr{ z^{2 \D_d}}{z_d^{2 \D_d}}  \lb c_{0,0} +  \sum_{j=1}^{\inf} c_{0,j} \ \ls \fr{z}{z_d} \rs^{2 j \D_d  }  
+ \sum_{i=1}^\inf  \sum_{j=1}^{\inf} c_{i,j} \  \ls \fr{z }{z_h } \rs^ {2 i}  \ls \fr{z}{z_d} \rs^{2 j \D_d }     \rb  , 
\ee
where condensation is again represented by $D = 1/z_d^{ \D_d}$ with the conformal dimension $\D_d$. Note that above the invariance under $z \to -z$, which changes only the overall sign of the action \eq{Action:scalar} and does not affect the equation of motion, allows the series expansion only with even powers. Comparing this expansion with the previous results, the coefficients of the leading correction are given by $b_{0,0} =-1$ and $c_{0,0} = - 1/4$.

Repeating the entanglement entropy calculation, we finally obtain the entanglement entropy in the UV regime with $l \ll z_h, \, z_d$ reads
\be
S_E = \fr{R}{2 G}  \ls \log \fr{l}{\e} + \fr{1}{24}  \fr{ l^2}{z_h^2}  -\frac{\sqrt{\pi }  \  \Gamma (\Delta_d +1)}{ 2^{4 + 2 \Delta_d }  \ \Gamma \left(\Delta_d +\frac{3}{2}\right)}   \ \fr{  l^{2 \Delta_d } }{z_d^{2 \Delta_d }} + \cdots \rs .
\ee
For $\D_d=1$, excitation and condensation have the same conformal dimension and the resulting entanglement entropy becomes
\be
S_E = \fr{R}{2 G}  \ls \log \fr{l}{\e} + \fr{\et}{24}  \fr{l^2}{\xi_{eff}^2}+ \cdots \rs ,
\ee
with an effective correlation length 
\be
\xi_{eff} = { \fr{ \sqrt {2} \, z_h  z_d } {  \sqrt {\left|  2 z_d^2 -  z_h^2 \right| }}  } ,
\ee
where $\et \equiv {\rm sign} (2 z_d^2 -  z_h^2)$. Due to the existence of $\et$,  the resulting entanglement entropy can show two different behaviors relying on the parameter range. For $2 z_d^2 > z_h^2$, excitation becomes dominant and increases the entanglement entropy. For $2 z_d^2 <  z_h^2$, oppositely, condensation is dominant and reduces the entanglement entropy.

For general $\D_d$ satisfying $1< \D_d<2$, the first correction changes its sign at the critical subsystem size $l_c$ given by 
\be
l_c^{2(\D_d -1)} =  \frac{2^{2 \Delta_d +1} \, \Gamma \left(\Delta_d +\frac{3}{2}\right) }{3 \sqrt{\pi } \, \Gamma (\Delta_d +1) } \fr{z_d^{2 \Delta_d }}{z_h^2}  .
\ee
This result implies that for $l < l_c$ excitation is dominant and that the entanglement entropy of the deformed theory is larger than that of the undeformed CFT. For $l > l_c$, on the other hand, condensation becomes dominant and reduces the entanglement entropy.  In the field theory deformed by excitation and condensation, the equal time two-point function of a local operator with a conformal dimension $\D_d$ is holographically given by
\be 
&& \bra O (x_1) \, O (x_2)\ket  \nn
&&\sim  \fr{1}{|x_1-x_2|^{2 \D_O}} \ls 1  - \fr{\D_O}{12}  \fr{ |x_1-x_2|^2}{z_h^2}  +  \frac{\sqrt{\pi }  \D_O \  \Gamma (\Delta_d +1)}{ 2^{3 + 2 \Delta_d }  \ \Gamma \left(\Delta_d +\frac{3}{2}\right)}   \ \fr{  |x_1-x_2|^{2 \Delta_d } }{z_d^{2 \Delta_d }} + \cdots \rs .
\ee
Excitation and condensation affect the two-point function in the opposite way to the entanglement entropy case. This is because, if excitation and condensation increases the entanglement entropy of the background theory, the two-point function of two local operators becomes weaker in the background having strong macroscopic correlation.

%
%

\section{Discussion}
\label{sec:7}

In this paper, we holographically studied the finite-size effects on the macroscopic entanglement entropy and the microscopic two-point functions. In this case, the finite-size effect is associated with the RG flow effect because the inverse of the finite size corresponds to the energy scale of the real space RG flow.  As a setup, we considered two-dimensional field theories deformed by excitation and condensation, which have dual gravitational descriptions, BTZ black hole and asymptotic AdS geometry deformed by the backreaction of the bulk scalar field, respectively. The macroscopic and microscopic correlations of the deformed two-dimensional QFT can be described by a geometrical object extending to the above three-dimensional dual geometries.

A relevant deformation, in general, triggers a nontrivial RG flow. At the critical UV and IR fixed points, in particular, the system has an infinite correlation length because of restoring of the scale symmetry. In the intermediate energy regime, however, the relevant deformation breaks the scale symmetry and makes the correlation length finite. Therefore, it would be important to clarify the relation between the correlation length and the deformations. We explicitly showed how the correlation length depends on the deformations in the UV and IR regions by exploiting the holographic propositions. First, we investigated the deformation effect on the entanglement entropy. The entanglement entropy measures the quantum correlation between two macroscopic subsystems so that we can regard the entanglement entropy as the macroscopic quantum correlation. We showed that the excitation (condensation) increases (decreases) the entanglement entropy in the UV region. Recalling that the entanglement entropy is associated with the $c$-function, which measures the degrees of freedom of the system, the change of the entanglement entropy caused by the deformations looks natural. The reason is that the excitation (condensation) increases (decreases) the degrees of freedom of the system. 

We also investigated the microscopic two-point function of local operators. To do so, we introduced two local operators to the deformed QFT. In this case, the deformed theory is regarded as the background medium containing excitation and condensation. As a result, the two-point function we considered corresponds to the microscopic quantum correlation of two local operators in the medium. For a two-dimensional QFT, intriguingly, the holographic dual of such a two-point function is again described by the same geometric curve like the one used in the entanglement entropy calculation. This fact may give a hint to the relation between the macroscopic and microscopic quantum correlations. The holographic result of the two-point function shows that, as opposed to the entanglement entropy case,  the excitation (condensation) decreases (increases) the two-point function of the local operator. This result is understood in the following way. The excitation, as mentioned before, increases the entanglement entropy of the background medium, which indicates that the ingredient of the background medium is correlated more strongly to each other. As the quantum correlation becomes strong, the background medium gives rise to a strong screening effect and the resulting two-point function of the local operator becomes weaker. For the condensation case, the entanglement entropy and the two-point function result in the opposite situation, as expected. When a system has both excitation and condensation, these two deformations compete with each other to increase or decrease the entanglement entropy. We also showed that, when the deformation operator is relevant, the excitation effect is dominant at the higher energy scale. On the contrary, the condensation effect becomes dominant at the lower energy scale.

When we calculated the two-point function, we used the formula proposed in Ref. \cite{Susskind:aa}, which maps the geodesic length in the dual geometry to the two-point function of the dual field theory. To check the validity of this proposal, we investigated the two-point function of the dual field theory by applying the holographic renormalization technique. We checked that the holographic renormalization leads to the same qualitative behavior as the one obtained by the proposal in Ref. \cite{Susskind:aa}.

\appendix
\section{Appendix} \label{sec:9}

Here we present details of the derivation of \eqref{slbtb} in Section \ref{sec:41}.
Inserting the normalized Green function \eqref{NGreen} into \eqref{Result:htwopoint} and performing the integration, we have 
\bea \label{NGF}
G(z,0;x_1^\m,x_2^\m)=- \frac{i\, e^{i\pi\Delta} z^\Delta }{2^{\Delta +1}\Gamma(\Delta)}\biggl(\frac{\sqrt{(2-\Delta)\Delta}}{z_h\sqrt{z^2+\vert x_1^\m-x_2^\m\vert^2}}\biggr)^\Delta 
H_{-\Delta}^{(2)}\biggl(\frac{\sqrt{(2-\Delta)\Delta}\sqrt{z^2+\vert x_1^\m-x_2^\m\vert^2}}{z_h}\biggr),
\nonumber \\
\eea
where $\Delta = \Delta_O$ and $H_n^{(2)}$ indicates the Hankel function of the second kind. Here we used the following integral formula
\bea 
\int_0^\infty &dk&k^{\nu+1}J_\nu(k \vert x_1^\m-x_2^\m\vert)K_\rho(z\sqrt{k^2-\alpha^2})(k^2-\alpha^2)^{-\rho/2}
\nonumber
\\
&=&\frac{\pi}{2}e^{-i\pi(\rho-\nu-1/2)}\frac{\vert x_1^\m-x_2^\m\vert^\nu}{z^\rho}\biggl(\frac{\sqrt{z^2+\vert x_1^\m-x_2^\m\vert^2}}{\alpha}\biggr)^{\rho-\nu-1}H_{\rho-\nu-1}^{(2)}(\alpha\sqrt{z^2+\vert x_1^\m-x_2^\m\vert^2}).
\nonumber
\\
\eea
Next, we want to compute the equal-time correlation function. To do so, it is useful to rewrite \eqref{NGF} in terms of an integral form as 
\bea
G(z,0;x_1^\m,x_2^\m)&=&z^\Delta
\frac{e^{i\pi\Delta/2}(\sqrt{(2-\Delta)\Delta})^{\Delta}}{2^\Delta\Gamma(\Delta)\vert x_1^\m-x_2^\m\vert^\Delta z_h^\Delta}\int^\infty_\infty dy \,e^{\Delta y-i \frac{\sqrt{(2-\Delta)\Delta}\,\vert x_1^\m-x_2^\m\vert\cosh y}{z_h}},
\eea
where $\vert x_1^\m-x_2^\m\vert=\sqrt{(t_1-t_2)^2-(x_1-x_2)^2}$. When we consider the equal-time correlation, $t_1=t_2$, this Lorentz invariant interval becomes space-like $\vert x_1^\m-x_2^\m\vert\rightarrow-i\vert x_1-x_2\vert$. Then, we have
\bea
G(z,0;x_1,x_2)&=&z^\Delta\frac{e^{i\pi\Delta/2}(\sqrt{(2-\Delta)\Delta})^{\Delta}}{2^{\Delta}\Gamma(\Delta)(-i\vert x_1-x_2\vert)^\Delta z_h^\Delta}\int^\infty_\infty dy \,e^{\Delta y-\frac{\sqrt{(2-\Delta)\Delta}\,\vert x_1-x_2\vert\cosh y}{z_h}}
\nonumber
\\
&=&z^\Delta\frac{e^{i\pi\Delta/2}(\sqrt{(2-\Delta)\Delta})^{\Delta}}{2^{\Delta-1}\Gamma(\Delta)(-i\vert x_1-x_2\vert)^\Delta z_h^\Delta}K_{-\Delta}\biggl(\frac{\sqrt{(2-\Delta)\Delta}\,\vert x_1-x_2\vert}{z_h}\biggr),
\eea
where we applied the following relation to the second line 
\bea
K_{-\sigma}(\alpha)\equiv\frac{1}{2}\int^\infty_\infty \,dy\,e^{\sigma y-\alpha \cosh y}.  .
\eea
Finally, the boundary-to-boundary two-point function is obtained by taking the limit $z\rightarrow 0$ together with the normalization factor $z^{-\Delta}$
\bea
G(x_1,x_2)=\lim_{z\rightarrow 0} z^{-\Delta} G(z,0;x_1,x_2).
\eea

\acknowledgments

We are grateful to Yunseok Seo for helpful discussions. C. Park is supported by Mid-career Researcher Program through the National Research Foundation (NRF) of Korea grant No. NRF-2019R1A2C1006639. J. H. Lee is supported by Basic Science Research Program through the National Research Foundation (NRF) of Korea funded by the Ministry of Education grant No. NRF-2018R1A6A3A11049655.


\vspace{0cm}

	\bibliographystyle{apsrev4-1}

\bibliography{References}

\end{document}